\documentclass{ptapap}

\author{Emese Plachy}[plachy.emese@csfk.mta.hu, CSFK]

\affil[CSFK]{Konkoly Observatory, Research Centre for Astronomy and Earth Sciences, Hungarian Academy of Sciences, H-1121 Budapest}

\title{Cepheid investigations in the era of space photometric missions}

\begin{document}

\maketitle

\begin{abstract}
Cepheid stars are crucial objects for a variety of topics that range from stellar pulsation and the evolution of intermediate-mass stars to the understanding the structure of the Galaxy and the Universe through the distance measurements they provide. The developments in hydrodynamical calculations, the release of large ground-based surveys, and the advent of continuous, space-based photometry revealed many puzzling phenomena about these stars in the last few years. In this paper I collected some important and new results in the topics of distance measurements and binarity investigations. I also summarize the most recent discoveries in their light variations, such as period doubling, modulation, low-amplitude additional modes, period jitter and the signs of granulation, and discuss the new opportunities that current and future space missions will offer for us.

\end{abstract}

\section{Introduction}

The most well-known property of Cepheids stars is that they are primary distance indicators. Our estimates about the size of Universe is based on our knowledge of Cepheids. The uncertainty of the Hubble constant is presently 2.4 percent based on extragalactic Cepheid distances in the Local Group \citep{riess}. Nowadays, the unprecedentedly accurate and continuous space photometric measurements reveal even the tiniest changes in the light variation, leading to a new question: are Cepheids really stable pulsators, and if not, are they standard candles yet?

Classical Cepheids are young intermediate-mass and metal rich stars that are useful tracers for the young disk populations, spiral arms and open clusters, while Type II Cepheids are old, low mass, metal poor stars placed in disk populations, globular clusters, the halo and the bulge. The latter type is divided into subtypes according to evolutionary stages and period ranges: BL Her (P<$\sim$4 d), W Vir ($\sim$4< P<$\sim$20 d) and ) RV Tau (P>$\sim$20 d) stars. There is roughly 1.5 magnitude difference in the absolute brightness between the two populations, but both types have strict period-luminosity relations. A third class is formed from those Cepheid-like stars that do not fit into either type therefore they are called anomalous Cepheids: they are 1-2 solar mass, metal poor, core helium burning stars.

\section{New distance measurements}

The period-luminosity (PL) relation, also called Leavitt’s law, honouring its discoverer, was found even before it was generally accepted that Cepheids, in fact, pulsate \citep{leavitt}. Many achievements have been made to find the definitive form of the law since then. The new measurements are taken in long wavelengths or by employing empirically defined Wesenheit indices \citep{madore} to minimize reddening effects. The relation is clearly metallicity dependent, therefore differs for different galaxies. The Magellanic Clouds play the most important role in the calibration, being the closest neighbour galaxies. They have been mapped in detail by OGLE Survey that discovered almost ten thousand classical Cepheids in them \citep{oglecef}. A possible break in, and nonlinearity of the PL relation were suggested \citep{pl} and are still debated. Effects of blends, bumps, eclipses and period changes on Leavitt’s law was also considered recently by \citet{valera}. 

Accurate and reliable distance measurements of Cepheids are essential to calibrate Leawitt's law: the first Gaia data release (DR1) provided new trigonometric parallaxes of 212 Cepheids \citep{gaiapar}. The radial pulsation property is also used  as a distance measurement technique known as the Baade-Wesselink (BW) method: if we measure the change in the angular diameter and the change in radius we are able to derive the distance of the object. The accuracy of this method is limited by the projection factor (p-factor) that converts radial velocity (RV) to pulsation velocity. We do not know the exact value of p-factor yet and its period dependence is also a matter of debate \citep{ngeow,breitfelder}. While the p-factor gives the largest uncertainty in the BW method, \citet{rvmod} pointed out another problem: if the RV and the photometric measurements are not simultaneous, the cycle-to-cycle changes that were detected in the RV curve of several stars, can cause significant systematic uncertainty. It is commonly believed that Cepheids are very regular variables, but this is not true for all Cepheids. The fraction of irregularly behaving Cepheids and the typical magnitude of the irregularities is unknown yet, only future statistics from space-based photometry can answer this.

Distance measurements can be very smart in peculiar cases. RS Puppis is the unique Cepheid star located in a reflection nebula. The light echoes of its brightness variation were used to determine the geometric distance of the nebula with polarimetric imaging taken by the Hubble Space Telescope \citep{kervella}. New distance measurements were published recently aimed at the region behind the bulge: 3 classical Cepheids were found to be in the nuclear bulge with approximately the same age \citep{matsunaga}, 5 other are located behind the Galactic bulge in the flared outer disk \citep{feast}, and 37 in the inner part of the bulge mostly behind the center of the Galaxy \citep{dekany}. Based on a different estimate for the extinction \citet{debate} debate this latter result, claiming that they are behind the inner disk. According to another interesting discovery 4 Cepheids were placed in a new dwarf galaxy behind the bulge at a distance of 90 kpc from the Galactic center \citep{dwarf}. This was shortly disproved when it turned out that 3 of these stars were measured by OGLE Survey and were found to be constant or spotted stars \citep{nodwarf}. The misclassification problem is indeed a serious issue, space photometric measurements also suggest that Cepheid impostors may reach high numbers \citep{keplercef,poretti2}. Therefore Cepheid candidates require careful analysis before using them as distance indicators.

\section{Binarity}

60\% of classical Cepheids are expected to have a companion \citep{binaries}, but because of their large size they cannot reside in close binaries. The variations of the RV or the $\gamma$-velocity are useful tracers of binarity. Recently 9 new spectroscopic binaries were found among southern Cepheids \citep{szabados1,szabados2}. The prototype $\delta$ Cephei was also suspected to be in a binary system \citep{deltacep}. Cepheids in eclipsing binary systems provide accurate radii and masses, however, given the large separation it is not easy to find eclipses. Fortunately a statistical amount of decade-long photometric measurements are available in the OGLE database, and several binary systems with a  Cepheid companion have already been published \citep{udalski}. The combined pulsating-eclipsing light curves provide not only physical properties with excellent accuracy but a distance-independent determination of the p-factor too \citep{pilecki}. We also know an exotic binary system (OGLE-LMC-CEP-1718) consisting of two first overtone classical Cepheid variables \citep{gieren}. The investigation of the OGLE-LMC562.05.9009 double-lined binary revealed two very similar stars in the system: both of them fall in the theoretical instability strip, but only one is pulsating \citep{strangebinary}. An underestimation of the temperature or metallicity differences was suggested as possible source of this strange result. Direct imaging of Cepheid companions is possible by now, 39 such system were resolved recently by snapshots of Hubble Space Telescope \citep{evans}. Long-baseline interferometry is also a powerful tool to detect close faint companions \citep{gallenne}.

\section{Pulsation characteristics}

The pulsation of Cepheids has been studied in detail since it was generally accepted that their light variation is governed by pulsation driven by the $\kappa$-$\gamma$ mechanism. Their characteristic pulsation amplitudes range from 0.2 to 2 magnitudes and the periods span from about one day to months. Cepheids are typical instances of radial stellar pulsation, which produce different features in the brightness variation according to the number of nodal surfaces in the star.
Their number of 0,1,2,3 correspond to the fundamental mode, first-, second-, and third-overtone pulsation, respectively. Type II Cepheids and most of the classical Cepheids are fundamental-mode pulsators. The mode selection was investigated in detail with one-dimensional hydrodynamical models: the different pulsation modes group in well-determined regions in the log$T_{\mathrm{eff}}$-log$L$ plane as it is shown in Figure 3 in \citet{buchler}. 
The majority of classical Cepheids are single-mode pulsators, but double- and triple-mode Cepheids also exist. Double-mode Cepheids, also referred to as beat Cepheids, are more frequent than triple-mode ones, from which we know only a few. The exact period ratios of the different modes are metallicity dependent, therefore it is useful information in metallic content studies \citep{beat}. One such study was recently published for the Galactic beat Cepheids \citep{beatmetal}. 

The fundamental-mode classical Cepheid light curves typically show a fast rise to the maximum and a slower decline. The shape of the descending branch is period dependent. In the 6-16 days period range we see the sign of the shock waves reaching the surface in the form of a secondary wave in the light curve, these are called bump Cepheids. The phase of the bump depends on the pulsation period, this property was christened Hertzsprung Progression after its discoverer \citep{hertzsprung}. Another subgroup, that contains mostly overtone-mode pulsators and called s-Cepheids, show symmetric light curves, that makes them easy to confuse with ellipsoidal binaries. 

The identification of anomalous Cepheids in the Galaxy is not unambigous either. Their period range overlaps with RR Lyraes, BL Her stars and the and short period classical Cepheids as well. However, a large number of anomalous Cepheids were investigated in the Magellanic Clouds with the OGLE survey and it was found that their difference in the light curve shapes can be used to distinguish them. By plotting the Fourier parameters $\phi_{21}$ and $\phi_{31}$ against the period they form well-defined groups \citep{2015anomalous}. 

The cycle-to-cycle variation in light curves is more common in Type II Cepheids than in the classical ones. The irregularities become more prominent toward the longer periods. A very characteristic feature, that almost all RV Tau stars clearly display, is the alternation of the small- and the large-amplitude cycles. 

\subsection{Nonlinear phenomena}

A central question in Cepheid pulsation is the role of nonlinear dynamics. Theoretical models from the 90's already predicted period doubling (PD) and chaos caused by half-integer resonances between pulsation modes for both Cepheid populations. With the novel, improved nonlinear convective hydrodynamical models the BL Her regime could be mapped and different routes to chaos have been found,  as well as periodic windows between the chaotic regions \citep{smolecmodel1}. The extension of this model grid shows that there are two domains where PD can occur, in a low luminosity region where the period is between 2 and 6.5 days, and a high luminosity range above 9.5 day period in the W Vir regime \citep{smolecmodel2}. These calculations also show the most probable resonances responsible for the PD. 
 
PD manifests as alternating cycles, and it was already identified in few BL Her stars in the OGLE data \citep{blpd2,blpd}. A W Vir-type was found to show this phenomenon in the extended mission of the \textit{Kepler} Space Telescope (Plachy et al. in prep.). In the latter case the subharmonic frequencies appeared in the Fourier spectra, as a clear sign of PD. PD can be the explanation of the alternating cycles found by \citet{templeton} in the light curve of the prototype, W Virginis itself. The other population II classical pulsating stars, the RR Lyraes, also undergo PD episodes \citep{pdrrl}, actually this discovery attracted the attention to this phenomenon again.

The already mentioned alternation of the minima in RV Tau-type brightness variation, that is the main characteristic of this type, is believed to be also caused by nonlinear effects and it is referred to PD. Low-dimensional chaos was detected in two RV Tau stars \citep{kiss,kollath}. The occasional interchanges in the order of the deep and shallow minima, which is seen in some RV Tau light curves \citep{interchange}, is typical in chaotic systems and support the theory that nonlinear behaviour is indeed what we observe. Similar interchanges are also visible in the period doubled cycles of RR Lyrae stars.

Although the PD in RR Lyrae stars seem to be accompanied with the Blazhko effect, which turned out to be a common feature in RR Lyraes, similar amplitude and/or phase modulation is very rare in Cepheids. The only modulated Galactic Cepheid that we know so far is V473 Lyrae in which two modulation periods were identified  \citep{V473}. We know other examples in the Magellanic Clouds from the OGLE Survey \citep{modcef,modcef2}.
The modulated Cepheids are overtone or double-mode pulsators and all have short periods.

\subsection{Nonradial modes}  

Thanks to the precision of OGLE, small-amplitude additional modes were recently discovered in Cepheid light curves, which form three groups near the 0.62 period ratio in the Petersen-diagram, just as the small-amplitude frequencies of RR Lyrae stars, resolved by space-based observations, do \citep{nonradial}. These additional periodicities were puzzling for a while as they could not be theoretically explained with radial or with nonradial pulsations either, therefore they have been called ``the mysterious 0.62 ratio" or $P_x$ and $F_x$ for the corresponding frequency. Now the solution seems to be found: due to surface cancellations the true frequency is not the observed $F_x$, that usually appears with the largest amplitude among the additional modes, but the half of it. With these considerations these modes can be modelled with linear nonadiabatic calculations and have been identified as nonradial modes with angular degrees from $\ell$ = 7 to 9 \citep{dziembowski}. Indeed, the 0.5 subharmonic of $F_x$ is present in a significant fraction of the observations.  

\section{Space photometry} 

Continuous and high precision space photometry is available for several Cepheid stars. The first space observation of a Cepheid was taken about Polaris using two instruments (the Solar Mass Ejection Imager on the \textit{Coriolis} spacecraft and the star tracker of Wide Field Infrared Explorer) that were not designed to do photometry. But the light curves turned out to be so useful that they led to two important discoveries. One is that the long since observed amplitude decrease of Polaris turned into increase suggesting its cyclic nature rather than monotonic. The other discovery was a low-amplitude additional variation on the time scale of 2-6 days that was concluded to caused by granulation \citep{bruntt}. 

The Canadian \textit{MOST} satellite observed four Cepheid stars. \citet{most} found the fundamental mode RT Aur to be more stable than the first overtone SZ Tau that showed fluctuations in the amplitude. The analysis of the MOST data of V473 Lyr and U TrA is in progress (Molnár et al. in prep., in these proceedings).

In the \textit{CoRoT} measurements 7 classical Cepheids and a BL Her-type star were identified \citep{poretti1,poretti2}. These stars do not show low-amplitude additional modes and only a very weak cycle-to-cycle variation was detected. However, a unique triple mode star was discovered among them, with an unusual period ratio.

The original \textit{Kepler} field contained only one classical Cepheid: V1154 Cyg. The first 600 days of data already showed a light curve with period jitter \citep{derekas1}, but the analysis of the whole 4 year data revealed a $\sim$158 day period amplitude modulation too \citep{kanev}. With the precision of \textit{Kepler}, solar-like oscillation can be easily detected, but this star does not show such oscillations, which is consistent with the theory that large-amplitude pulsation blocks solar-like oscillations. However, granulation was detected with the time scale of 3-5 days \citep{derekas2}. \textit{Kepler} provided the first space photometry of an RV Tau-type star (DF Cyg) as well \citep{bodi}.

The K2 mission of the \textit{Kepler} space telescope provides a unique opportunity for the astronomical community to propose targets for high-quality photometric measurements: hundreds of Cepheids have been already proposed and observed during the first ten campaigns \citep{targetselection}. The future launch of \textit{TESS}  mission promise us the greatest number of space observation about Cepheids ever.
Many details are expected to emerge about the pulsation from the space-based photometry that may change our picture about these magnificent stars.

\acknowledgements{This project has been supported by the Hungarian NKFIH Grants PD-121203 and K-115709. Useful comments of L.\ Moln\'ar and L.\ Szabados are acknowledged.}

\bibliographystyle{ptapap}
\bibliography{eplachy_brite2}

\end{document}